\documentclass[lettersize,journal]{IEEEtran}
\IEEEoverridecommandlockouts

\usepackage{amsmath,amssymb,amsfonts}
\usepackage{algorithmic}
\usepackage{graphicx}
\usepackage{textcomp}
\usepackage{xcolor}
\usepackage[style=numeric,natbib=true]{biblatex} 
\graphicspath{{./Images/}} 
\usepackage{mathtools}
\usepackage{amsmath}
\usepackage{fancyhdr} 
 
\def\BibTeX{{\rm B\kern-.05em{\sc i\kern-.025em b}\kern-.08em
    T\kern-.1667em\lower.7ex\hbox{E}\kern-.125emX}}
\begin{document}

\title{Design of an Efficient Three-Level Buck-Boost Converter in PSIM} 
\author{\IEEEauthorblockN{Justin London \\}
\IEEEauthorblockA{\textit{Dept. of Electrical Engineering and Computer Science} \\
\textit{University of North Dakota}\\
Grand Forks, North Dakota \\
justin.london@und.edu}}

\maketitle

\begin{abstract}
    Compared to conventional converters, a three-level buck-boost (3L-BB) converter offers higher efficiency, reduced switching losses, and increased power density.  We design a 3L-BB converter given certain voltage and current specifications in PSIM.  We simulate the circuit in PSIM and analyze the power, voltage, and current waveforms by comparing the observed simulated values in PSIM with their mathematically driven theoretical values.  We examine its power efficiencies and determine if the circuit meets given DC distribution specifications.  We show that the proposed three-phase design, which uses two DC-DC single-ended primary-conductor converters (SEPICs), is power efficient and is a compelling solution for high-power and high-voltage applications. 
\end{abstract}


\section{Introduction}
    Buck-boost converters are power converters that can either increase or decrease input voltages to produce a regulated DC output voltage.  Given their versatility, they ensure stable output voltage for optimal performance even when the battery voltage falls below a target level. Energy harvesting systems, such as solar panels and thermoelectric generators, have variable output voltages due to environmental conditions.  They can handle a wide range of input voltages, which is beneficial in battery-powered applications where the voltage can significantly drop as the battery discharges.  
    
    In comparison to other types of converters, they can achieve high power efficiency (typically 85\%-95\%) when designed properly, making it suitable for energy-sensitive applications.  Buck-booster converters can be used to stabilize voltage \cite{Suryadi:2020}. They maintain a consistent output voltage, even when the input voltage fluctuates, ensuring stable operation of connected devices.  By adjusting the duty cycle of the internal switch, buck-boost converters can maintain a relatively constant output voltage even when the input voltage varies. Thus,  buck-boost converters can achieve high efficiency, minimizing power loss during voltage conversion, especially when operated within their optimal range. They can handle a broad range of input voltages, making them adaptable to various power sources like batteries, AC-DC adapters, or renewable energy systems. 

    Three-level buck-boost converters are primarily used to efficiently step up or step down voltage in various applications like portable electronics, energy harvesting, automotive systems, and industrial equipment, while maintaining a stable output voltage and minimizing losses. These converters offer advantages such as increased efficiency, higher power density, and reduced component stress compared to traditional two-level converters.  Various topologies of three-phase AD-DC converters have been proposed \cite{Rivera:2018, Falin:2021} 

    Multi-level buck-boost converters offer several advantages over traditional single-level converters, particularly in high-power and high-voltage applications. These advantages include lower voltage stress on power semiconductor devices, reduced output voltage ripple, and the ability to achieve higher voltage gains with fewer components. Additionally, they can handle higher power levels and offer better efficiency and harmonic characteristics.  Multi-level converters can achieve higher voltage conversion ratios (both step-up and step-down) compared to single-level converters, especially when using a large number of levels. This can be achieved without resorting to transformers or extreme duty cycles.  This capability is crucial in applications requiring a wide range of input and output voltages. Multi-level converters are inherently capable of handling higher power levels due to the distributed voltage across multiple devices. This makes them suitable for demanding applications like industrial motor drives and electric vehicles.
    
    Multi-level converters can achieve higher voltage conversion ratios (both step-up and step-down) compared to single-level converters, especially when using a large number of levels. This can be achieved without resorting to transformers or extreme duty cycles. 
    This capability is crucial in applications requiring a wide range of input and output voltages. 
      
    We seek to design a three-stage DC-DC buck-boost converter and to analyze whether the design meets given voltage and current specifications and the power efficiency.  In our design, two DC-DC single-ended primary-conductor converter (SEPICs) are used.  PSIM is used to simulate the designed three-stage cascaded buck boost converter.  The results are discussed and analyzed followed by a conclusion.

\section{Topologies}

    A traditional buck switching converter consists of two metal-oxide silicon field-effect transistors (MOSFETs), one inductor, an input in parallel with the input source and an output capacitor \cite{Falin:2021}.   The switch gate-drive signals are complementary running at duty cycles D and 1-D.  To distribute power more efficiently and enable smaller footprints, a three-phase design can be used.   In a two-phase converter, the two power stages operate $180^{\circ}$ out of phase with one another. A three-phase converter adds a third-stage operating $120^{\circ}$ out of phase with the others.  In addition, it provides improved ripple cancellation creating a much smoother output voltage since the current ripples from each phase are staggered, canceling each other out more effectively at the output.   Consequently, with less ripples to filter out, the size and weight of the output inductors and capacitors can be substantially reduced.   
    
    There are various three-phase topologies.  The typical architecture consists of three phase-modular AC/DC converters.  Each module can operate from the AC grid or a common DC link, converting the AC voltage to a DC voltage and vice versa.  In a Y-inverter architecture, three identical buck-boost converter modules are connected in a Y or star configuration to common star point \cite{Antivachis:2018}. This design allows for continuous AC voltage waveforms and accommodates a wide range of DC input or output voltages.  Menzi et al \cite{Menzi:2021} propose a six-module Y-rectifier topology and analyze the required modulation and control methods for single-phase and three-phase operation.  
    
\begin{figure}[h]
	   \centering
	   \includegraphics[width=0.7\linewidth]{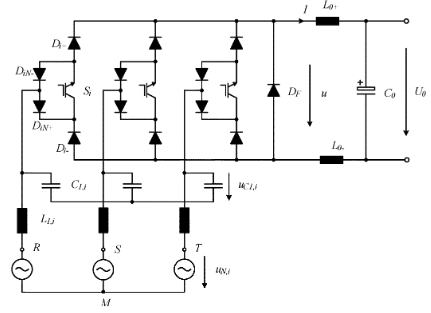}
	   \caption{Three phase three-switch buck rectifier. Source: 
       \cite{Nussbaumer:2007}}  
        \label{fig:three} 
    \end{figure}

    In a three-phase bridge PWM rectifier with series switch architecture, a three-phase AC/DC buck-boost converter includes a three-phase bridge PWM rectifier \cite{Malesani:1987, Wang:2021, Nussbaumer:2007} followed by an additional series switch on the DC side.  The switch controls the flow of current through a DC-side inductor allowing the converter to adjust the DC voltage and power flow.  The topology of a three phase three-switch buck-type rectifier is shown in Figure \ref{fig:three}.
    
\begin{figure}[h]
	   \centering
	   \includegraphics[width=0.7\linewidth]{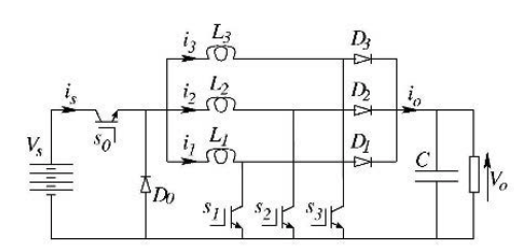}
	   \caption{Three phase interleaved converter. Source: 
       \cite{Alajmi:2022}}  
        \label{fig:interleave} 
    \end{figure}

    In an interleaved architecture, a non-inverting interleaved arrange of buck-boost modules reduce output current ripple, improve power output, and enhance power efficiency \cite{Alajmi:2022}. With interleaving converters, two or more converters are connected in parallel, which can increase the converter rated output power as well as output voltage.  The coupled inductor based interleaving converter can operate in continuous mode conduction (CCM) or discontinuous conduction mode (DCM).  When the interleaved converter operates in DCM model, it has a higher number of operating modes during one switching cycle compared to CCM operation.  Since DCM can reduce the passive component size and weight of the converter, DCM operation is preferred for electric vehicle (EV) and photovoltaic (PV) applications \cite{Ray:2010, Barry:2015}.  The ripple frequency of a three-phase interleaved converter is triple that of the switching frequency.   Alajmi et al. propose a three-phase interleaved converter is shown in Figure \ref{fig:interleave}.   Their topology consists of a single buck converter cascaded with $n$ parallel-interleaved boost converters which has a reduced switch count compared to the interleaved parallel cascaded non-inverting buck-boost converter (CNIBBC) provided in 

    Melo et al. \cite{Melo:2020} propose a single-phase equivalent circuit and space state model from non-conventional three-phase inverters based on a bidirectional DC-DC buck boost topology using isolated gate bipolar transistors (IGBT).  They show that the state space dynamic behavior variables of single-phase and three-phase models are similar and therefore an equivalent single-phase circuit can be used to represent each leg of a three-phase buck-boost inverter. 
\section{Design}
\begin{figure}[h]
	\centering
	\includegraphics[width=0.55\linewidth]{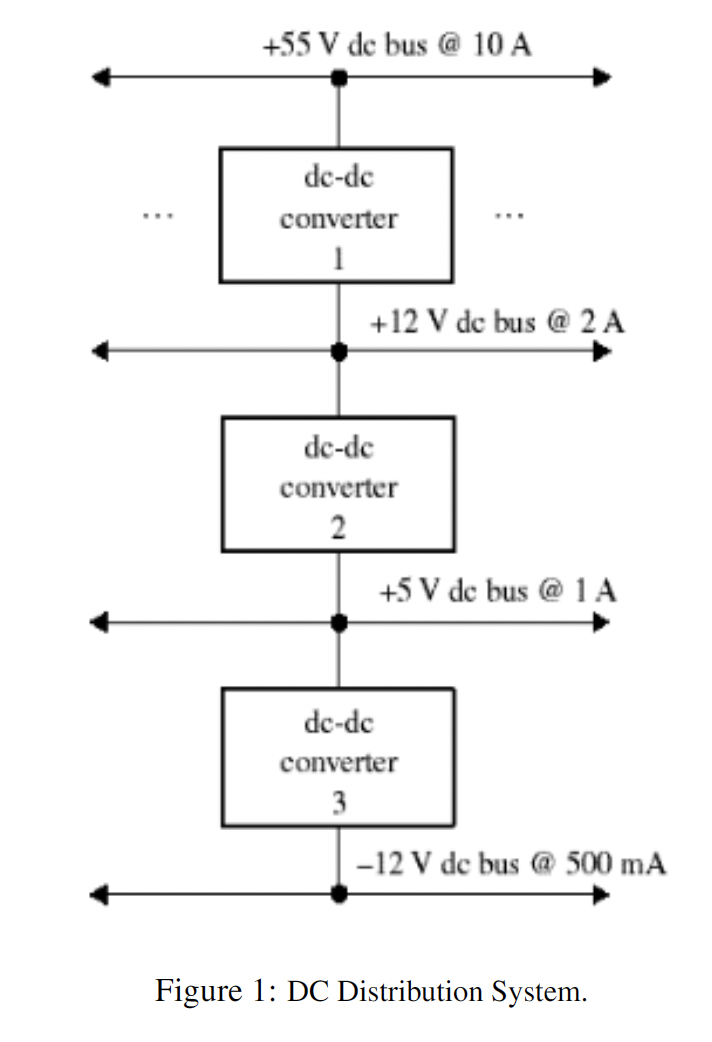}
	\caption{Design Specifications}  
    \label{fig:design} 
\end{figure}

    The three-stage buck-booster converter is designed to meet the following specifications as shown in Figure \ref{fig:design}.
    Since the source voltage of the first stage is $V_{s} = +55V @ 10A$ and the output voltage of the first stage is $V_{o} =  +12V @ 2A$ is positive and there is no polarity reversal, a single-ended primary inductance converter (SEPIC) is used as the first converter.  SEPICs have the advantage that they have short-circuit resistance and a non-inverted output, but have more complex designs that the standard buck-boost converter. 
    
    In a SEPIC, there are two inductors and the currents in the are constant.  There are two capacitors that are both large and the voltages across them are constant.  The circuit is assumed to operate in the steady state, meaning that voltage and current waveforms are periodic \cite{Hart:2011}.   Give a duty ratio $D$, the switch is closed for time $DT$ and open for $(1-D)T$.  The switch and the diode are assumed to be ideal.  $T = 1/f = 10 \mu s$.   
    
    Average voltage across $C_{1} = V_{s} - V_{o} = 55 - 12 = 43 V$, so that if we assume the ripple voltage across $C_{1}$ is no more than 0.5\%, then $V_{C_1} = 43 \times 0.005 = 0.215 V$.  We assume that the output voltage ripple $\Delta V_{0}$ is no more than $1\%$.
    
\begin{figure}[h]
	\centering
	\includegraphics[width=1\linewidth]{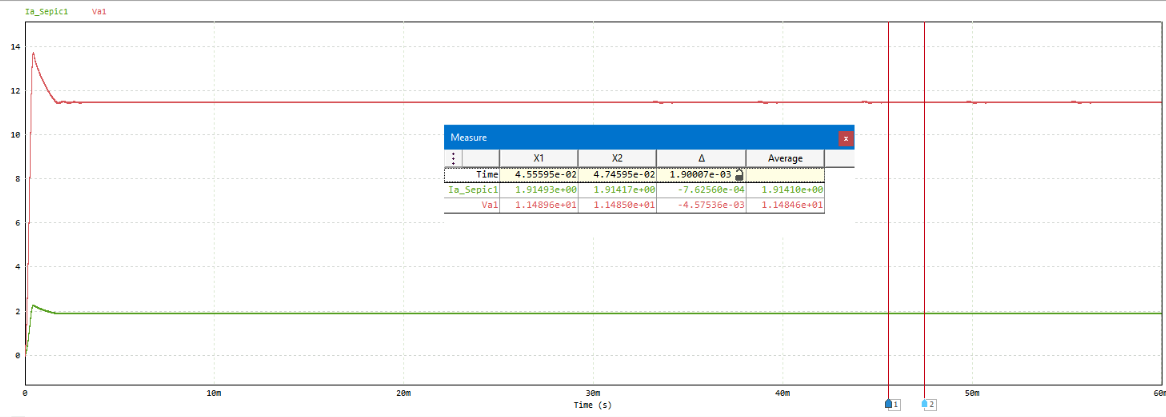}
	\caption{First-stage average voltage and current output}  
    \label{fig:output1} 
\end{figure}

    The output resistance is $V_{o}/I_{o} = 12 V/2 A = 6 \ \Omega$.  A switching frequency of $f = 100K$ for the first MOSFET switch $Q_{1}$.  We compute the duty cycle as 
    \begin{equation}
    D = \frac{V_{o}}{V_{o} + V_{s}} = 12/67 = 0.179 
    \end{equation}
    The capacitance of the first capacitator $C_{1}$ is 
    \begin{equation}
        C_{1} = \frac{DV_{o}}{R\Delta V_{C_1}f} = (0.179*12)/(0.215*600K) = 16.67 \ \mu F
    \end{equation}
    The capacitance of the second capacitor is \begin{equation}
        C_{2} = \frac{D*V_{0}}{R \Delta V_{o}f} = (0.179*12)/(0.01*600K) = 358 \ \mu F.  
    \end{equation}
    We assume continuous current in the inductors so that average current must be greater than one-half the change in current.  The minimum inductor sizes for continuous current are 
    \begin{equation}
        L_{1,min} = \frac{(1-D)^2R}{2Df} =  112.97 \ \mu H.
    \end{equation}  
    and 
    \begin{equation}
        L_{2,min} = \frac{(1-D)R}{2f} = (0.821*6)/200k = 24.6 \ \mu H.
    \end{equation}  
    To ensure continuous current operation, we select an inductor that is $25\%$ larger than $L_{min}$ so that $L_{1} = 141.21 \mu H$ and $L_2 = 30.8 \mu H$.  Figure \ref{fig:converter1} shows the first-stage SEPIC converter:
\begin{figure}[h]
	\centering
	\includegraphics[width=1\linewidth]{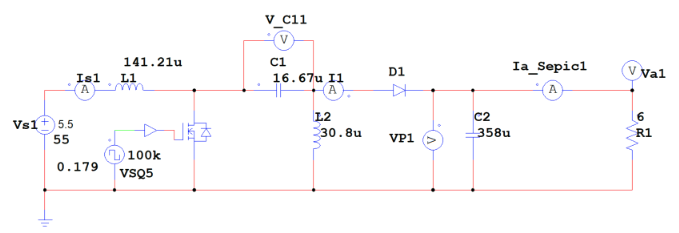}
    \caption{First-stage SEPIC converter}
    \label{fig:converter1} 
\end{figure} 
    The PSIM output (average) voltage and output current is shown in Figure \ref{fig:output1}.
    As shown the average output voltage is 11.5 V and the average output current is 1.91.  There is an error of $4.17\%$ and $4.5\%$, in the calculate output and current voltages, respectively, compared to their specified voltage of 12 V and 2 A, respectively, due to power loss across the MOSFET switch. 

    The second converter is also chosen to be a SEPIC since the input source voltage is $V_{s2} = +12V$ (the output voltage from the first converter) and the second output voltage is $V_{o2} = +5V$ so that there is no reverse in polarity.  The resistance is $V_{o2}/I_{o2} = 5 V/1 A = 5 \Omega$.  
    Average voltage across $C_{21}$ is +12 - 5 = +7V.  We assume the switching frequency $f = 100K \ Hz$ and that the ripple voltage across $C_{21}$ is no more than $1\%$, so that $\Delta V_{C_{21}} = 7 \cdot 0.01 = 0.07$.  Using the equations in (1) through (5) and using the same assumptions to maintain continuous current, we find
    \begin{align}
        D_{2} &= 5/17 = 0.294 \\
        C_{21} &= (0.294 \cdot 5)/(5 \cdot 0.07*100K) = 42 \  \mu F \\
        C_{22} &= (0.294 \cdot 5)/(5 \cdot 0.01*100K) = 294 \ \mu F \\
        L_{21,min} &= (0.706^2 \cdot 5)/(0.294 \cdot 200K) = 42.38 \ \mu H \\    
        L_{22,min} &= (0.706 \cdot 5)/200K = 17.65 
        \end{align}
        We increase $L_{21,min}$ and $L_{22,min}$ by $25\%$ so that $L_{21} = 52.98 \ \mu H$ and $L_{22} = 22.1 \ \mu H$.   

\begin{figure}[h]
	\centering
	\includegraphics[width=1\linewidth]{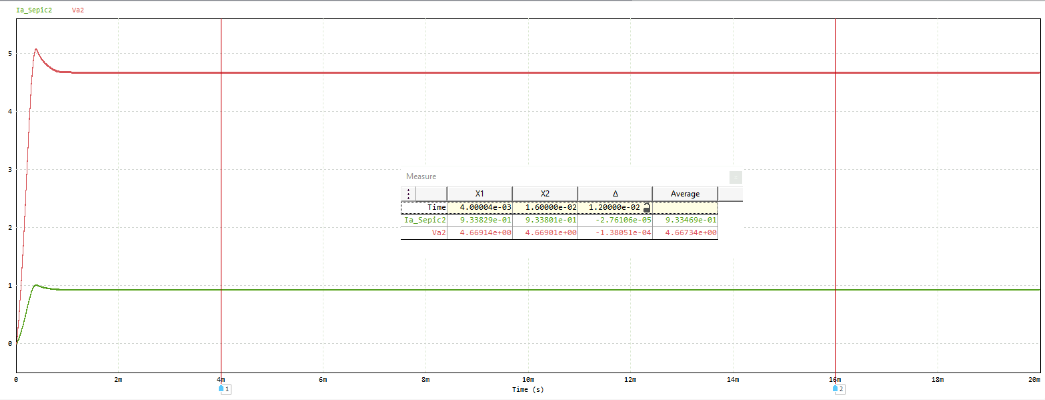}
    \caption{Average voltage and current output of second converter (run separately)}
    \label{fig:conv2} 
\end{figure}

\begin{figure}[h]
	\centering
	\includegraphics[width=1\linewidth]{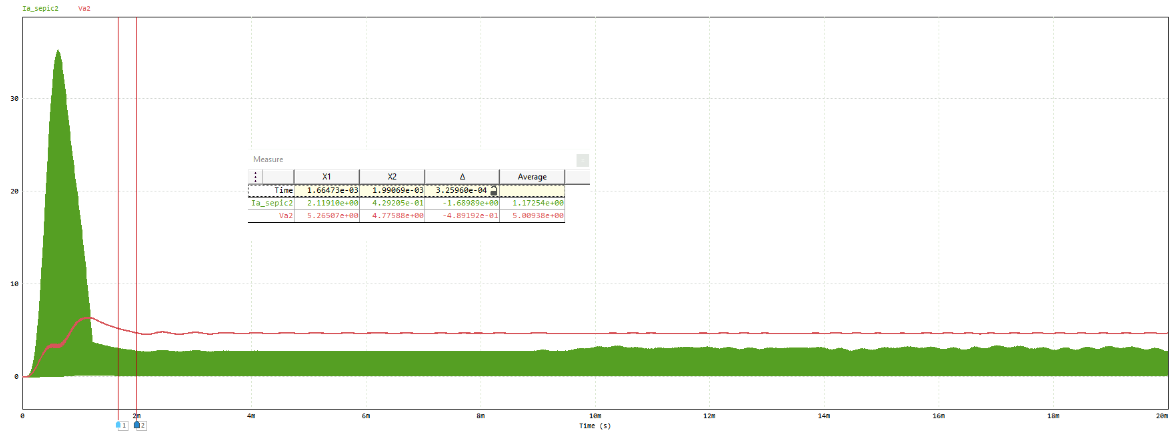}
    \caption{Average voltage and current output of second converter}
    \label{fig:output2} 
\end{figure}

\begin{figure}[h]
	\centering
	\includegraphics[width=0.9\linewidth]{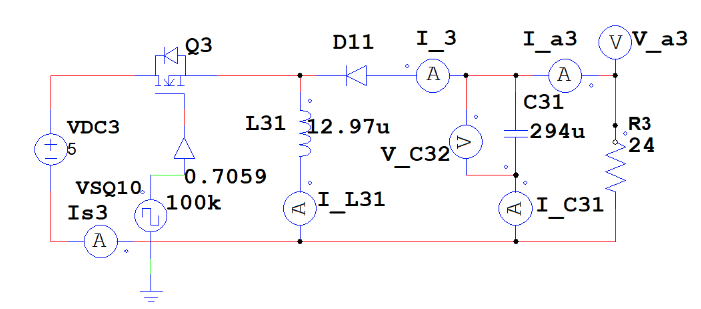}
    \caption{Third-stage buck-boost converter}
    \label{fig:converter4} 
\end{figure}

        Using Kirchhoff's current law, the diode and switch currents are:
        \begin{align}
        i_{D_{1}} &=
        \begin{cases}
            0 \ \ \ \ \ \ \ \ \ \ \ \ \ \ \text{when switch is closed} \\
            i_{L_{11}} + i_{L_{12}} \ \ \ \text{when switch is open} 
        \end{cases} \\
        i_{sw} &=
        \begin{cases}
            i_{L_{11}} + i_{L_{12}} \ \ \  \text{when switch is closed} \\ 
            0 \ \ \ \ \ \ \ \ \ \ \ \ \ \ \  \text{when switch is open}  
        \end{cases}
        \end{align}
        Using the average PSIM simulated values from Figure \ref{fig:numbers1}, $i_{D_{1}} = i_{L_{11}} + i_{L_{12}} = 0.581 - 2.62 = -2.04 A$ when the switch is open and $i_{sw} = -2.04 A$ when the switch is closed.  
        The second-stage converter is shown in Figure \ref{fig:design2}.  
\begin{figure}[h]
	\centering
	\includegraphics[width=0.9\linewidth]{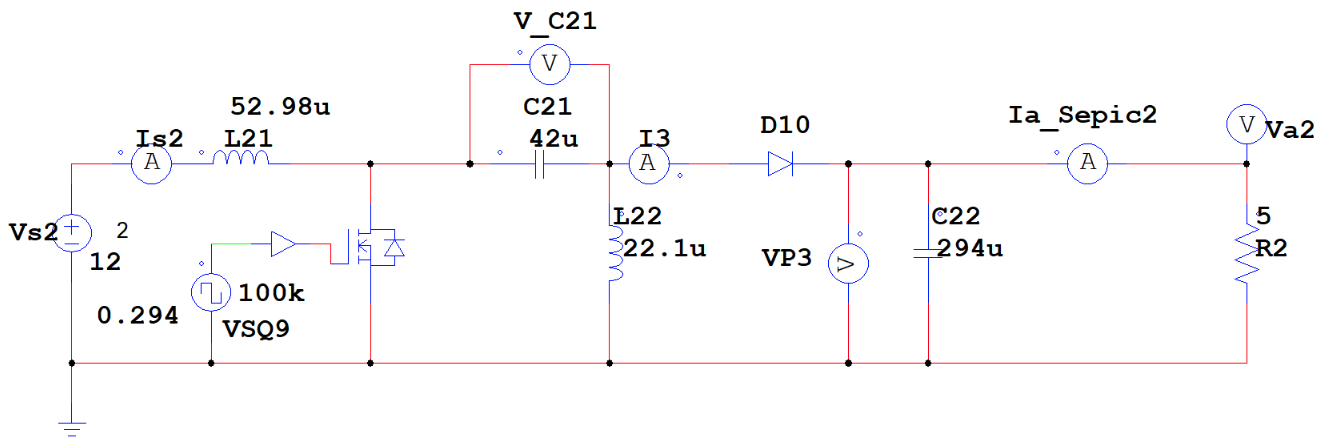}
    \caption{Second-stage converter}
    \label{fig:design2} 
\end{figure} 
    When the second converter is run separately in PSIM, we get the average output voltage and current of 4.67 V and 0.933 A, respectively, as shown in Figure \ref{fig:conv2} which are close to the specifications, but not exact with simulate measurement errors of $6.6\%$ and $6.7\%$, respectively.
    As shown in Figure \ref{fig:output2}, when the second SEPIC converter is connected into the three-phase buck-booster converter circuit, the output voltage of the second converter is +5 V and the average output current is 1.17 which is a difference of 0.17 A.  However, we note that the average output current ripple is higher than when the converter is run separately.
    Using the average PSIM simulated values from Figure \ref{fig:numbers2}, $i_{D_{2}} = i_{L_{21}} + i_{L_{22}} = 0.739 -1.92A$ = -1.18 A when the switch is open and $i_{sw_2} = -1.18A$ when the switch is closed.
    The third-stage converter selected is a standard buck-boost converter as there is polarity reversal since the input source voltage is $V_{s_3} =  +5 V$ and the output voltage is $V_{o_3}$ = -12 V.  The duty cycle is 
    \begin{equation}
    D_{3} = \frac{\lvert V_{o} \rvert}{\lvert V_{o} \rvert + V_{s_3}} = 0.7059
    \end{equation}
    The switching frequency is set to $f_3 = 100K \ Hz$ so that $T_3 = 10 \ \mu s$. The resistor value is $R_3 = \lvert -12 V \rvert/0.5 A = 24 \Omega$. The minimum inductor value is 
    \begin{equation}
        L_{3,min} = \frac{(1-D_{3})^2R_{3}}{2f_3} = \frac{(0.294^2 \cdot 24)}{200K} = 10.4 \ \mu H
    \end{equation}
    We assume the output ripple voltage is $\frac{\Delta V_{o_3}}{V_{o_3}} = 1\%$.   The capacitance is calculated as:
    \begin{equation}
        C_{31} = \frac{D_{3}}{R_{3}(\frac{\Delta V_{o_3}}{V_{o_3}})f_{3}} = \frac{0.706}{(24 \cdot 0.01 \cdot 100K)} = 29.4 \ \mu F
    \end{equation}
    Maximum and minimum inductor currents are determined as:
    \begin{equation}
        I_{max} = I_{L} + \frac{\Delta i_{L}}{2} = \frac{V_{s_3}D_{3}}{R_{3}(1-D_{3})^2} + \frac{V_{s_3}D_{3}T_{3}}{2L_{31}}
        \label{eq:Imax}
    \end{equation} 
        and 
    \begin{equation}
       I_{min} = I_{L} - \frac{\Delta i_{L}}{2} = \frac{V_{s_3}D_{3}}{R_{3}(1-D_{3})^2} - \frac{V_{s_3}D_{3}T_{3}}{2L_{31}}
       \label{eq:Imin}
    \end{equation}
    yielding $I_{max} = 1.75A$ and $I_{min} = 1.2A$.
    The third-stage buck-boost converter is shown in Figure \ref{fig:converter4}. 
The average output voltage and average output current matches the specifications of -12 V and 0.5 A, respectively.
\begin{figure}[h!]
	\centering
	\includegraphics[width=0.8\columnwidth]{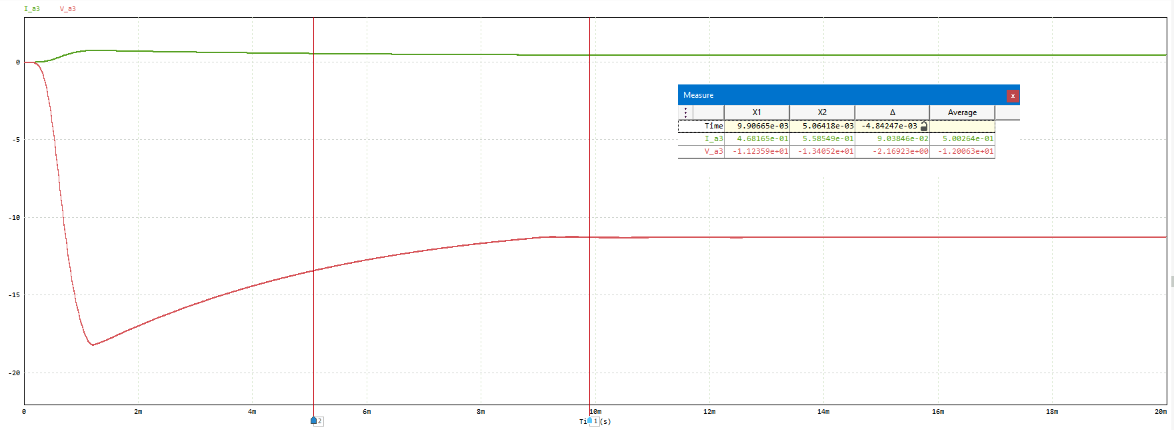}
    \caption{Average, Voltage and current output of third-stage converter}
    \label{fig:converter3} 
\end{figure} 

\section{Designed Three-Stage Converter}
\begin{figure}[h!]
	\centering
	\includegraphics[width=1\linewidth]{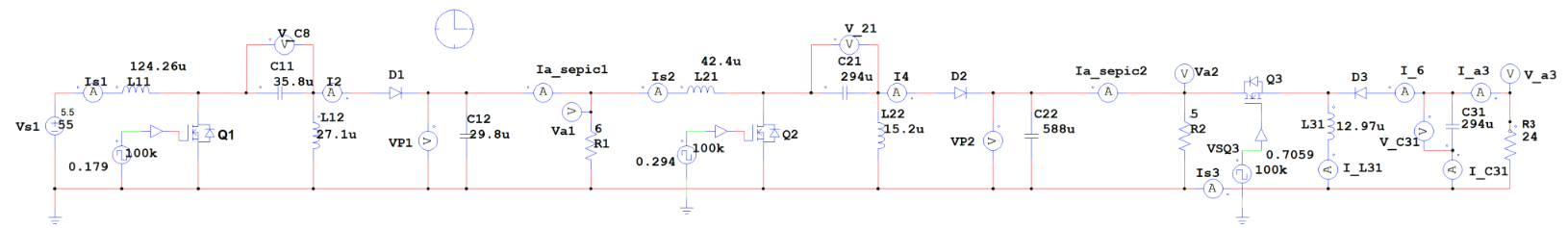}
    \caption{Three-stage converter designer}
    \label{design2} 
\end{figure}

\begin{figure}[h!]
	\centering
	\includegraphics[width=0.9\linewidth]{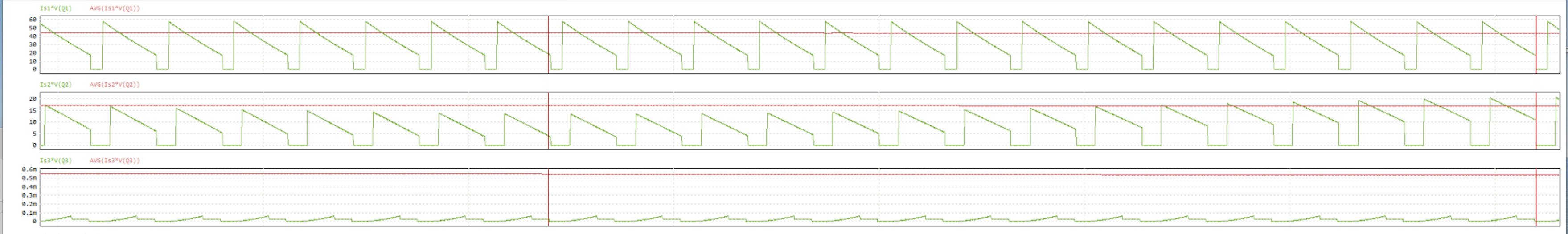}
    \caption{Power Waveforms for converters}
    \label{fig:waveforms} 
\end{figure}

\begin{figure}[h!]
	\centering
	\includegraphics[width=0.9\linewidth]{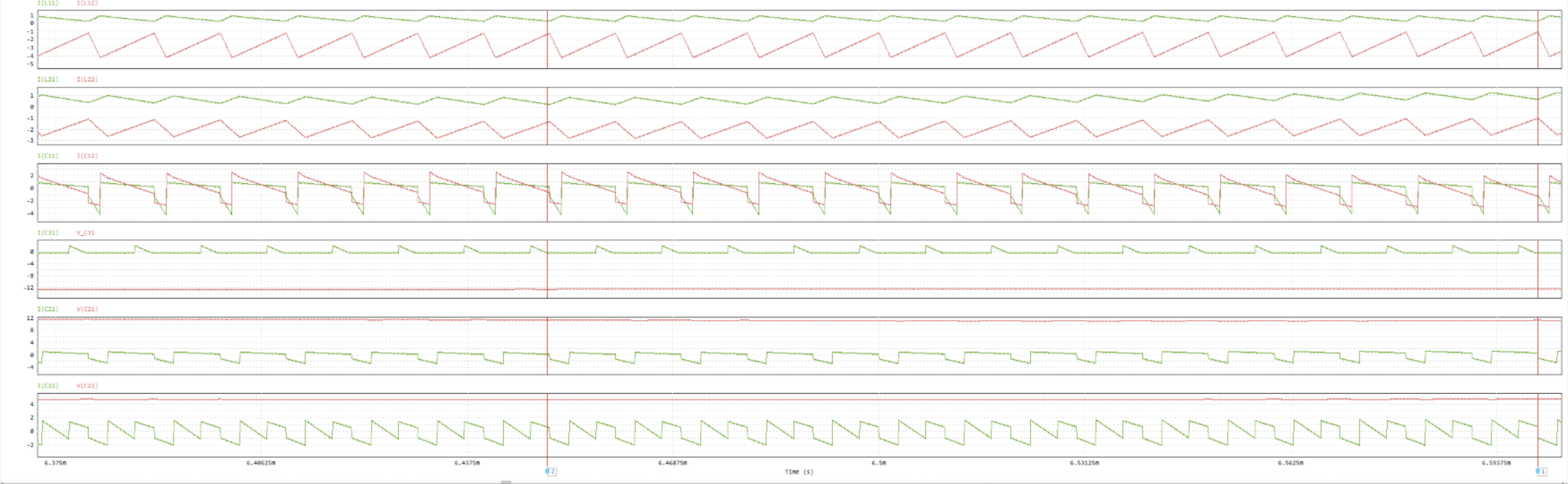}
    \caption{Waveforms of Capacitors and Inductors}
    \label{fig:numbers1} 
\end{figure}

\begin{figure}[h!]
	\centering
	\includegraphics[width=0.9\linewidth]{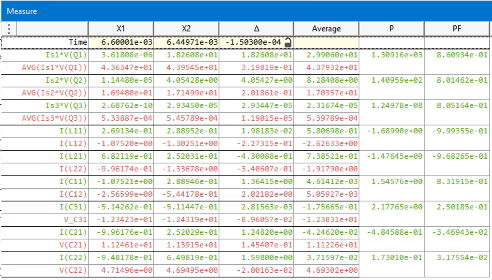}
    \caption{Power, voltage, and inductor measurements}
    \label{fig:numbers2} 
\end{figure}

    Figure \ref{design2} shows the complete three-stage buck-boost converter.  The input into the second and third converter are the voltage outputs from the first and second converters across the resistance loads, respectively.
   Figure \ref{fig:waveforms} shows plots of currents and voltage waveforms for various converter components.  The first plot shows average power across the MOSFET switch in the first converter.  The average power across the switch is 43.8 W and the power factor is $0.86$.  The second plot the average power across the MOSFET switch in the second converter.  The average power across the switch is 17.0 W and the power factor is $0.8$.  The third plot plots the average power across the MOSFET switch in the third converter.  The average power across the switch is $54 \ \mu W$ and the power factor is .
    Figure \ref{fig:numbers1} show the PSIM waveforms of the currents and voltages for the capacitors and inductors for the three converters.  The average capacitor and inductor voltages and current values as well as the apparent power and power factor are shown in Figure \ref{fig:numbers2}.
    The simulated peak-to-peak ripple of currents through the inductors and capacitors for each of the converters is shown in Figure \ref{fig:numbers2}.
\section{Power Dissipation}
    The main causes of power dissipation in a DC-DC converter are inductor conduction losses, power dissipated in the MOSFET, power loss in the capacitor filter, and diode conduction loss \cite{Tan:2006}. 
    Assuming the three-stage converter is ideal such that there are no energy dissipated in any components and switches, the power efficiencies are calculated as followed. The first converter has $V_{s_1} = 55$ and $I_{s_1} = 10 A$ so that $P_{s} = P_{in_1} = 55 \cdot 10 = 550 W$.  $P_{o_3}  = V_{o}I_{o} = 12 \cdot 2 = 24 W.$  The power efficiency of the first converter is $\eta_1 = P_{out_1}/{P_{in_1}} = 24/550 = 0.0436$.  The second converter has a power efficiency of $\eta_{2} = P_{out,2}/P_{in,2} = 5/24 = 0.208.$  The third converter has a power efficiency of $\eta_{3} = P_{out,3}/P_{in,3} = \lvert -24 \rvert/5 = 4.8$. However, power efficiency cannot be negative or greater than 1.  Thus, $\eta_{3}$ becomes 1.  
    In the non-ideal case, the output power efficiency of each converter can be calculated as:
    \begin{equation}
        \eta = P_{o}/(P_{o} + P_{total \ losses})         
    \end{equation}  
    where $P_{total \ losses} = P_{L} + P_{Q} + P_{C} + P_{D}$ + other losses where $P_{L}$ are inductor losses, $P_{Q}$ are the power losses from the switch, $P_{C}$ are power losses from the capicitator, and $P_{D}$ are the power losses from the diode \cite{Vorperian:2010}.  We analyze the three-stage converter in the non-ideal case.
    \subsection{Power Dissipated in the Inductor}
    In the third converter, the peak-to-peak of the ripple current through the the inductor current $I_{L_31}$ is equal to the sum of the DC input current $I_{s_3}$ and the DC output current $I_{o_3}$.
    The peak value of the switch current is 
    \begin{equation}
        I_{L_{31}} = \frac{V_{s}D_{3}}{R(1-D_{3})^2} = \frac{5*(0.706)}{24(0.294)^2} = 1.70 A. 
    \end{equation}
    The peak-to-peak variation in capacitor current is the same as the maximum inductor current \cite{Hart:2011} $I_{L_31} = I_{max} = 1.75A$ in Eq. \ref{eq:Imax}.  The current difference over the capacitor is  $I_{s} - I_{o} = 1 - 0.5 = 0.5$.  If we assume peak-to-peak inductor current does not exceed $10\%$ of the average value, then $\Delta i_{C} \approx \Delta i_{L} =  1.75 \cdot 0.1 = 0.175$  The allowable peak-to-peak output voltage ripple is $1\%$ so $-12 \cdot 0.01 = -0.12 V$.
    Therefore, the peak-to-peak output equivalent series resistance (ESR) is given by:
    \begin{align}
        \Delta V_{oESR} &= \Delta i_{C}r_{C} = I_{L,max}r_{C} \nonumber \\
                         &= I_{Lmax}\bigg ( \frac{\Delta V_{o_3}}{\Delta i_{C_31}} \bigg ) = 1.75 \bigg (\frac{-0.12}{0.175} \bigg ) = -1.2 V 
    \end{align}
    where $r_{C} = -0.686$.
    Let $r_{L}$ be the inductor equivalent series resistance.  Then $r_{L} = \Delta V_{L}/I_{L}$  The average inductor current, which is also the average source current is
    \begin{equation}
        I_{L_3} = I_{s} = \frac{V_{o}I_{o}}{V_{s}} = \frac{V^2_{o_3}}{V_{s}R_3} = \frac{(-12)^2}{5\cdot 24} = 1.2 A   
    \end{equation}
    Assume the voltage drop across the inductor is $1\%$ then $\Delta V_{L} = (5 - (-12) = 17 \cdot 0.01 = 0.17$ yielding $r_{L} = 0.17/1.2 = 0.142$
    The inductor conduction loss is \cite{Hairik:2019}
    \begin{equation}
        P_{L} = RI^2_{L,rms} = \frac{r_{L}I^2_{o}}{(1-D)^2} = \frac{0.142(0.5)^2}{(0.294)^2} =  0.411 \ W
    \end{equation}
\subsection{MOSFET Power Dissipated}
    Assume $r_{DS}$ to be the resistance of the MOSFET switch during the ON-state period.  The two main losses are conduction and switching losses.  To calculate the conduction loss, we assume that the inductor current $I_{L}$ is ripple-free and equal to the DC current $I_{s} + I_{o}$.  Therefore, the current of the MOSFET switch can be approximated as:
    \begin{align}
    i_{sw} = 
    \begin{cases}
        I_{L} = I_{s} + I_{o} = \frac{I_{o}}{1-D} = 2 A \ \ \ \text{for} \ 0 < t \leq DT \\
        0 \ \ \ \ \ \ \ \ \ \ \ \ \ \ \ \ \ \ \ \ \ \ \ \ \ \ \ \ \ \ \ \ \text{for} \ DT < t \leq T
    \end{cases}
    \end{align}
    The switch current RMS is
    \begin{align}
            I_{sw} &= \sqrt{\frac{1}{T} \int^T_{0} i^2_{sw}dt} = \frac{I_{o}}{(1-D)}\sqrt{\frac{1}{T}\int^{DT}_{0} dt} \\
            &= \frac{\sqrt{D}I_{o}}{(1-D)} = \frac{\sqrt{0.706} \cdot 0.5}{1-0.706} = 1.68 A.
    \end{align}
    The MOSFET conduction loss is
    \begin{equation}
        P_{R_{DS}} = R_{DS}I^2_{sw,rms} = \frac{D R_{DS} I^2_{o}}{(1-D)^2} = \frac{D R_{DS} P_{o}}{(1-D)^2R_{L}}
    \end{equation}
    Assuming that $R_{DS(on)} = R_{L} = 0.1 \Omega$, then $P_{R_{DS}} = \frac{0.706 \cdot 0.1 \cdot 0.5^2}{(0.249)^2} = 0.285 W$. 
   The switching losses can be assumed to remain constant with any changes in output voltages as they are assumed independent of output voltage.
\subsection{Filter Capacitor Power Loss}
\ \ \ The current of the output capacitor
\begin{equation}
    i_{C} = 
    \begin{cases}
        -I_{o} = -0.5 A \ \ \ \ \ \ \ \ \ \text{for} \ 0 < t \leq DT \\
        I_{s} = \frac{DI_{0}}{1 - D} = 1.42 A \ \ \ \text{for} \ DT < t \leq T.
    \end{cases}
\end{equation}
    So its RMS value will be 
\begin{equation}
    I_{Crms} = \sqrt{\frac{1}{T} \int^{T}_{0} I^2_{C}}dt = I_{0}\sqrt{\frac{D}{1-D}} = 2.84 V.
\end{equation}
The loss of power of the capacitor
\begin{equation}
    P_{C} = r_{C}I^2_{C_rms} = \frac{Dr_{c}I^2_{0}}{1-D} = \frac{0.706 \cdot -0.686 \cdot 0.5^2}{0.294} = -0.412 W
\end{equation}
\subsection{Diode Conduction Loss}
    The diode current can be approximated as
\begin{align}
    i_{D} = 
    \begin{cases}
        0 \ \ \ \ \ \ \ \ \ \ \ \ \ \ \ \ \ \ \ \ \ \ \ \  \text{for} \ 0 < t \leq DT \\
        I_{L} = I_{s} + I_{o} = I_{0} \ \ \ \ \text{for} \ DT < t \leq T
    \end{cases}
\end{align}
Th root mean squared error of the diode current is
\begin{align}
    I_{Drms} &= \sqrt{\frac{1}{T} \int^{T}_{0} i^2_{D}dt} = \frac{I_{0}}{1-D}\sqrt{\frac{1}{T}\int^{DT}_{0} dt} \\
    &= \frac{I_{0}}{\sqrt{1-D}} = \frac{0.5}{1-0.706} = 1.7 A.
\end{align}
    Denote $R_{f}$ as the feedback resistance and $R_{L}$ as the inductor resistance.   The power loss due to $R_{f}$ is
\begin{equation}
    P_{R_f} = R_{f}I^2_{D,rms} = \frac{R_{f}I^2_{0}}{1-D} 
\end{equation}
    The average value of $i_{D}$ is
\begin{equation}
    I_{D} = \frac{1}{T} \int^{T}_{0} i_{D}dt = \frac{I_{0}}{(1-D)T} \int^{T}_{DT}dt = I_{0} 
\end{equation}
The power loss of the forward voltage $V_{f}$, which in a typical diode is the minimum voltage required to allow current to flow through the diode in the forward direction, which we assume to be 0.6 V, is
\begin{equation}
    P_{V_f} = V_{f}I_{D} = V_{f}I_{o} = \frac{V_f P_{o}}{V_{o}}
\end{equation}
so that the total diode power loss is
\begin{align}
    P_{D} &= P_{V_f} + P_{R_f} = V_{f}I_{o} + \frac{R_{f}I^2_{0}}{1-D} \\
          &= \bigg (\frac{V_f}{V_{o}} + \frac{R_f}{(1-D)R_{L}} \bigg )P_{o}
\end{align}
    We assume $R_{f} = 0$, so that $P_{D} = V_{f}I_{o} = 0.6 \cdot 0.5  = 0.3 W$. 
    A similar analysis and calculations can be made for the first two SEPIC converters.  

\section{Power Efficiency Improvement}
    If we assume that the given voltage and current specifications are the capacity values, we get improved power efficiencies by using a lower input source power of 35 W as shown in Figure \ref{fig:design2} rather than the maximum capacity of 550 W.
    \begin{figure}[h]
	\centering
	\includegraphics[width=0.55\linewidth]{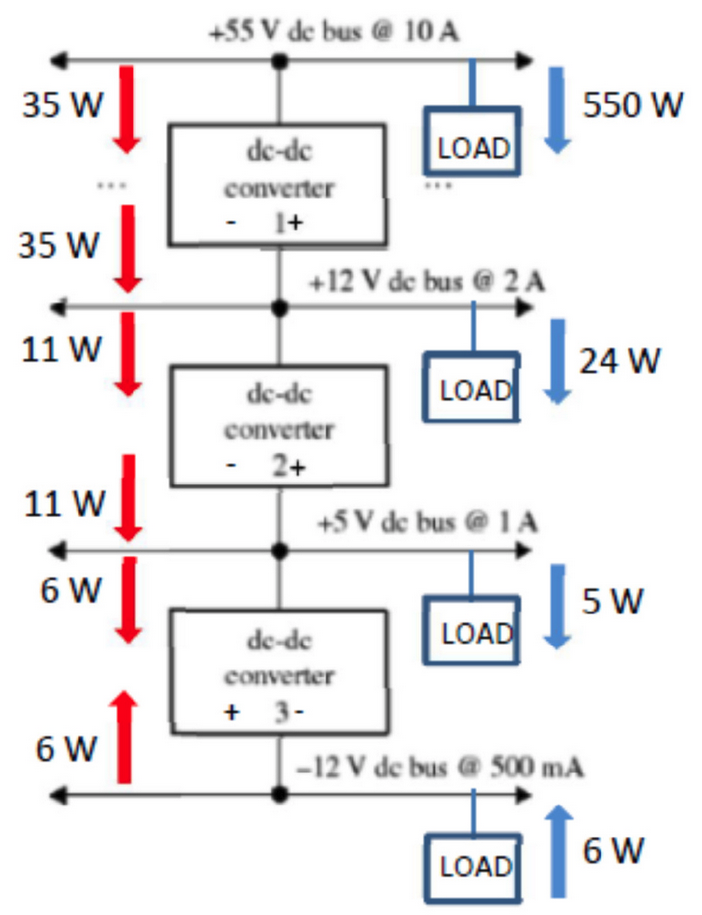}
	\caption{Design Specifications}  
    \label{fig:design2} 
\end{figure} 
For instance, for the first SEPIC converter, $P_{in} = V_{s_1}I_{s_1} = 35 W = 55 \cdot I_{s}$ yielding $I_{s_1} = 0.636 A.$  $I_{s_1} = V_{o_1}^2/(V_{s_1}R)$ or $0.636 = 12^2/(12R)$ so that $R_{1} = 4.12 \Omega.$.  We assume $f = 100K$ as before for each of the converters.  The power efficiency is 11/35 = 0.314.  Likewise, for the second SEPIC converter, $P_{in} = 11 W = 12 I_{s_2}$ yielding = 1.09 A and $R_{2} = 12/1.09 = 11 \Omega.$. The power efficiency is $\eta_{2} = 6/11 = 0.545$ The capacitor and inductor values can be computed from equations (2)-(5) for both SEPIC converters.  The duty cycles for both converters remain 0.179 and 0.294.  Finally for the standard buck boost converter, $P_{in} = 6 = 5I_{s_3}$ yielding $I_{s_3} = 1.2 A$ and $R_{3} = 4.16 \Omega$.  The output current is $I_{o_3} = 12/4.16 = 2.88 A$.  Thus, the power of the third converter is $P_{o_3} = 12 \cdot 2.88 = 34.6 W$. The power efficiency is $\eta_{3} = 34.6/6 = 5.77$ which becomes 1 since power efficiency cannot exceed 1.  The inductor and capacitor values for the third converter can be computed from equations (14) and (15).
\section{Conclusion}
    The simulated results in PSIM of the designed three-phase DC-DC buck-boost converter illustrate that it meets the design objectives.  The simulated waveforms for voltage and currents across the switches and other passive components are as expected.  In particular, there are sawtooth/periodic patterns in the current and voltages. In a non-ideal case, inefficiencies due to some power losses from power absorbed and dissipated by switches, diodes, inductors, and capacitors could account for why the first two converters are close, but do not exactly meet the specifications. The simulations were re-run numerous times by fine-tuning the capacitor and inductor values to try and get all the simulated voltage and current values to exactly match the specifications.  Though the values are not all exact but quite close, based on the voltage drop and output ripple assumptions used in the theoretically calculated inductor and capacitor values in each converter, the results are reasonable.  Overall, the first two SEPIC converters have low power efficiencies (by specification), while the third buck boost converter has very high power efficiency.  Therefore, the three-stage buck boost converter is efficiently designed.
    A three-stage buck-boost converter like the one designed offers a compelling solution for high-power and high-voltage applications due to its advantage in voltage stress reduction, voltage gain, efficiency, ripple reduction, and harmonic performance.
\printbibliography

@article{Suryadi:2020,
author = {Suryadi, A. and Asmoro, P. and Sofwan, A.},
title = {Design and Simulation Converter with Buck-boost Converter as the Voltage Stabilizer},
journal = {International Journal of Electrical, Energy and Power System Engineering},
year = {2020},
volume = {3},
number = {3},
pages = {77-81}
}

@book{Hart:2011,
author = {Hart, D.},
title = {Power Electronics},
publisher = {McGrawHill Higher Education},
year = {2011},
}

@article{Hairik:2019,
author = {Hairik, H. and AbdulAbass, A. and Abbas, K.},
title = {DC/DC Buck-Boost Converter Efficiency and Power Dissipation Calculating at Operating Points Not Included in the Datasheet},
journal = {Journal of Multidisciplinary Engineering Science and Technology},
volume = {6},
year = {2019}
}

@inproceedings{Tan:2006,
author = {Tan, S-C. and Lai, Y.M. and Tse, C. and Wu, C.K.},
title = {A PulseWidth Modulation Based Integral Sliding Mode Currnet Controller for Boost Converters},
booktitle = {37th IEEE Power Electronics Specialistics Conference},
year = {2006},
pages = {1612-1618}
}

@article{Vorperian:2010,
author = {Voperian, V.},
title = {Simple Efficiency Formula for Regulated DC-to-DC Converters},
journal = {IEEE Transactions on Aerospace and Electronic Systems},
volume = {46},
number = {4},
pages = {2123-2131},
year = {2010}
}

@inproceedings{Rivera:2018,
author = {Riveria, M. and Faundez, D. and Kolar, J. and Wheeler, P. and Riveros, J.},
title = {Three-Phase AC-DC Converters with Passive, Actiev and Hybrid Currnet Injeciton Circuits},
booktitle = {Proceedings on the 4th IEEE Argentina Biennial Congress},
year = {2018}
}

@article{Falin:2021,
author = {Falin, J .and Aguilar, A.},
title = {Maximize power density with three-level buck-switching chargers},
journal = {Analog Design Journal - Texas Instruments},
year = {2021}
}

@article{Alajmi:2022,
author = {Alajmi, B. and Marei, Mostafa and Abdelsalam, I. and Ahmed, N.},
title = {Multiphase Interleaved Converter Based on Cascaded Non-Interverting Buck-Boost Converter},
journal = {IEEE Access},
year = {2022}
}

@inproceedings{Antivachis:2018,
author = {Antivachis, M. and Bortis, D. and Schrittwieser, L. and Kolar, J.},
title = {Three-phase bucki-boost Y-inverter with wide DC input voltage range},
booktitle = {Annual IEEE Conference on Applied Power Electronics Conference and Exposition (APEC)},
year = {2018}
}

@masters{Wang:2021,
author = {Wang, Y.},
title = {Analysis of Three-phase Rectifier via Three Different Control Methods and Switch Power Loss Comparison},
school = {Minnesota State University, Mankato},
year = {2021}
}

@article{Nussbaumer:2007,
author = {Nussbaumer, T. and Baumann, M. and Kolar, J.},
title = {Comprehensive Design of a Three-Phase Three-Switch Buck-Type PWM Rectifier},
journal = {IEEE Transactions on Power Electronics},
year = {2007}
}

@article{Malesani:1987,
author = {Malesani, L and Tenti, P.},
title = {Three-phase AC/DC PWM converter with sinusoidal AC currents and minimum filter requirements},
journal = {IEEE Trans. Industrial Applicaitons},
volume = {IA-23},
number = {1},
pages = {71-77},
year = {1987}
}

@inproceedings{Ray:2010,
author = {Ray, B. and Koasi, H. and McNeal, S. and Jordan, B. and Scofield, J.},
title = {A comprehensive multi-mode performance analysis of interleaved boost converters},
booktitle = {Proceedings IEEE Energy Conversion Congress Exposition},
year = {2010},
pages = {3014-3021}
}

@article{Barry:2015,
author = {Barry, B. and Hayes, J. and Rylko, M.},
title = {CCM and DCM operation of the interleaved two-phase boost converter with discrete and coupled inductors},
journal = {IEEE Trans. Power Electronics},
volume = {30},
number = {12},
pages = {6551-6567},
year = {2015},
}

@article{Melo:2020,
author = {Melo, V. and Melo, A. and Santos, W. and Fardini, J. and Encarnacao, L.},
title = {Open-Loop Single-Phase Space State Model and Equivalent Circuit of a Non-Conventional Three-Phase Inverter},
journal = {MDPI Electronics},
year = {2020}
}

@inproceedings{Menzi:2021,
author = {Menzi, D., and Kolar, J. and Everts, J.},
title = {Single-Phase Full-Power Operable Three-Phase Buck-Boost Y-Rectifier Concepts},
booktitle = {Proceedings of the 36th Applied Power Electronics Conference and Exposition (APEC 2021)},
year = {2021}
}

\end{document}